# Orthogonal Sampling based Broad-Band Signal Generation with Low-Bandwidth Electronics

Mohamed I. Hosni, Janosch Meier, Younus Mandalawi, Karanveer Singh, Paulomi Mandal, Ahmed H. Elghandour and Thomas Schneider

*Abstract*—. **High-bandwidth signals are needed in many applications like radar, sensing, measurement and communications. Especially in optical networks, the sampling rate and analog bandwidth of digital-to-analog converters (DACs) is a bottleneck for further increasing data rates. To circumvent the sampling rate and bandwidth problem of electronic DACs, we demonstrate the generation of wide-band signals with low-bandwidth electronics. This generation is based on orthogonal sampling with sinc-pulse sequences in $N$ parallel branches. The method not only reduces the sampling rate and bandwidth, at the same time the effective number of bits (ENOB) is improved, dramatically reducing the requirements on the electronic signal processing. In proof of concept experiments the generation of analog signals, as well as Nyquist shaped and normal data will be shown. In simulations we investigate the performance of 60 GHz data generation by 20 and 12 GHz electronics. The method can easily be integrated together with already existing electronic DAC designs and would be of great interest for all high-bandwidth applications.**

## I. INTRODUCTION

With the exponential growth of communications as well as radar, lidar, sensing and other bandwidth hungry applications, the generation of high-quality, broad-band signals become increasingly important [1-3]. The required signal bandwidth of these applications can reach several tens or even hundreds of gigahertz [4]. For 6G and beyond for instance, peak data rates of 1 Tbit/s are foreseen [5]. Depending on the scale of parallelization and spectral efficiency of the modulation format and coding, this can require DAC with analog bandwidths of tenths of GHz. For electronic and especially integrated DAC it's quite challenging to meet these bandwidths demands. Cutting-edge CMOS-based DACs have shown sampling rates of 97 GS/s and bandwidths of 40 GHz using 7-nm FinFET technology [6]. Higher bandwidths and sampling rates are possible in SiGe BiCMOS or InP heterojunction bipolar transistors [7, 8] but the interconnection between the DACs and the following digital signal processing will be challenging. Additionally, with increased bandwidths, jitter and quantization noise problems will deteriorate the signal quality, which reduces the resolution as measured by the effective number of bits (ENOB) [9]. Therefore, recently approaches utilizing CMOS based DACs together with an external multiplexer have reached much attention. The multiplexing can be accomplished in the time [10-13] or frequency domain [14]. By time interleaving two sub-DACs, a 100 GS/s signal generation was presented in [10] and [11]. However, only the sampling rate can be increased by this method. The bandwidth of the generated signal is the same as that of the sub-DACs. A method that can enhance the sampling rate and bandwidth has been presented in [13]. By adding up the outputs of two sub-DACs in an analog multiplexer (AMUX), a bandwidth of 110 GHz was shown. However, the analog multiplexing needs a very sophisticated pre-processing and a very fast switching. Another approach is to frequency-interleave single low-bandwidth sub-spectra to a high-bandwidth signal [14]. The basic idea is quite simple, but the practical implementation requires a pre-processing of the data and a bank of filters with well-defined transmission curves.

In this paper, we present a novel and very simple approach for the multiplexing of $N$ low-bandwidth sub-DACs based on orthogonal sampling by sinc-pulse sequences. For ideal components, the sampling with sinc-pulse sequences is error-free. Additionally, the sinc pulse sequence generation is carried out with a radio frequency oscillator, which can show very low jitter values, and for the sampling a simple multiplication is used. Therefore, not only the bandwidth and sampling rate can be enhanced, at the same time the ENOB of the DAC system is increased. Since $N$ sub-DACs can be multiplexed simultaneously, the requirements on sampling rate and bandwidth of the single electronic DAC can be drastically reduced by increasing the number of parallel branches.

In proof of concept experiments we demonstrate the generation of analog as well as Nyquist shaped and normal data signals. By simulations we investigate the possibilities of the method for increasing the resolution of the generated signals. The proposed method is very simple and does not require electronic pre-processing or fast switching. For the multiplexing only simple radio frequency devices like an oscillator, a multiplier and an adder are needed. Therefore, the co-integration with already existing DAC designs seems to be straight-forward. Therefore, we believe that it will be of interest for all kind of applications were high-bandwidth signals are needed.

The paper is organized as follows; In the next section we will describe the basic principle of the method. In section III the experimental and simulation results are presented and Sec. IV concludes the paper.

Mohamed I. Hosni, Younus Mandalawi, Janosch Meier, Karanveer Singh, and Thomas Schneider are with the THz Photonics Group, Technische Universität Braunschweig, Schleinitzstraße 22, 38106 Braunschweig, Germany (e-mail: mohammed.elsayed@ihf.tu-bs.de).

Ahmed H. Elghandour is with the Communication department, Military technical college (MTC), Cairo 11766, Egypt.
Mohamed I. Hosni and Janosch Meier contributed equally to the paper.

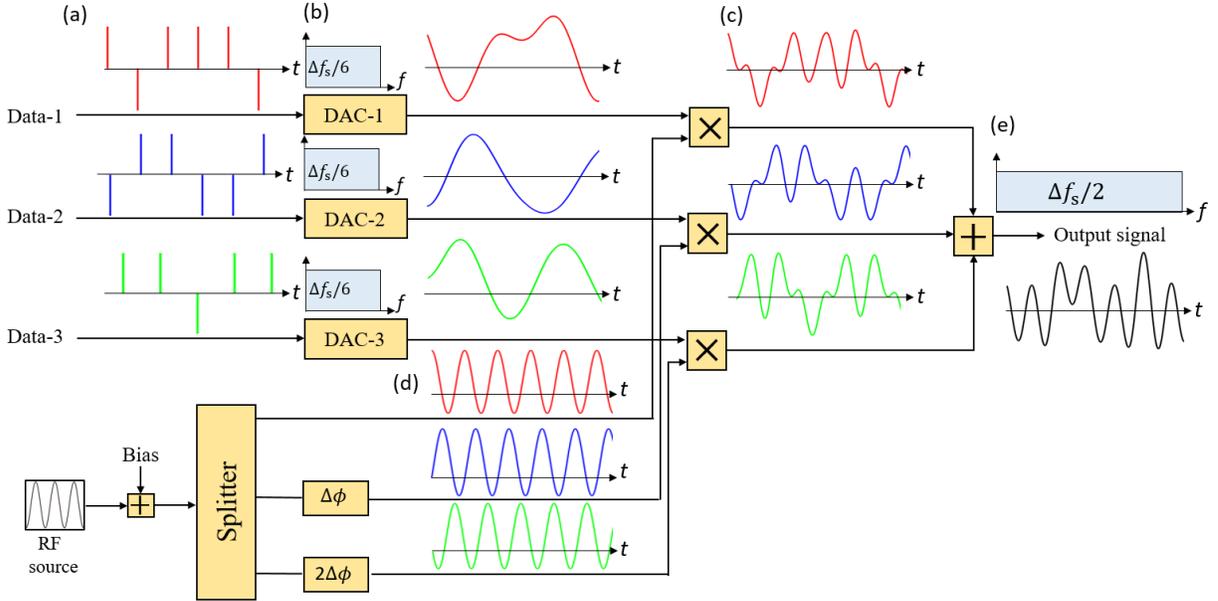

Fig. 1 Generation of a broad-bandwidth signal by orthogonal sampling. For the sake of simplicity, the number of branches is restricted to $N = 3$. In each of the branches, the low rate digital input (a) is converted to a low-bandwidth analog signal by a digital to analog converter (DAC) with the bandwidth $\Delta f_s/6$ (b). Afterwards, in each branch the low-bandwidth waveforms are sampled (multiplied) with orthogonal, sinc pulse sequences (c). The single sinc pulse sequences is just a single radio frequency (RF) with a constant offset [14] (d). For orthogonality, the sinc pulse sequences in the branches have to be time shifted against each other, so that the maxima of the sequence in the next branch are in the zero crossings of the previous. This can be achieved by a phase change of the RF by $\Delta \phi = 360°/3 = 120°$ between the branches (d). Finally, the time-shifted sequences, weighted with the sampling points are added up to form the high-bandwidth signal with a bandwidth of $\Delta f_s/2$ and a sampling rate of $\Delta f_s$ (e).

## II. PRINCIPLE OF OPERATION

An ideal DAC can be mathematically described by a Dirac delta sequence weighted with the sampling values and filtered by a rectangular filter function. The Dirac Delta sequence and the rectangular filter function are both unlimited in frequency and therefore cannot be achieved in a practical device. Thus, for a practical DAC the single pulse in the sequence and the transfer function of the filter are rather a mixing between a rectangular and a Gaussian function.

From the mathematical description of sampling it follows that no Dirac Delta sequence is necessary for an ideal DAC. Each signal limited to the bandwidth $\Delta f_s/2$ can be generated by the superposition of time-shifted, single sinc pulses, weighted with the corresponding sampling values [15]:

$$s(t) = \sum_{k=-\infty}^{\infty} s\left(\frac{k}{\Delta f_s}\right) \cdot \text{sinc}(\Delta f_s t - k). \qquad (1)$$

In complete contrast to a Dirac Delta, which is infinitesimal short, the sinc pulse is unlimited in time. However, if the time-shift between the sampling points is $1/\Delta f_s$, the sinc pulses are orthogonal to each other. Thus, there is no mutual influence between the sampling points and the sampling rate is $\Delta f_s$. But, since sinc pulses, are unlimited in time, they are also only a mathematical construct.

Alternatively, the signal can be seen as the superposition of time-shifted, orthogonal sinc-pulse sequences, weighted with periodic sampling points. A sinc pulse sequence is the unlimited superposition of time-shifted, single sinc pulses [16] and can be expressed as:

$$\text{sq}_{N,\Delta f_s}(t) = \sum_{k=-\infty}^{\infty} \text{sinc}(\Delta f_s t - kN)$$
$$= \frac{2}{N}\left(\frac{1}{2} + \sum_{k=1}^{\frac{N-1}{2}} \cos\left(\frac{2\pi k\Delta f_s t}{N}\right)\right) \qquad (2)$$

Although the single sinc pulse is unlimited in time, the orthogonality between the time shifted pulses ensures the zero inter-symbol-interference between the pulses in the superposition. Between two pulses in the sequence there are $N-1$ zero crossings. As single sinc pulses, sinc pulse sequences are also orthogonal to each other, if the next sequence is time shifted to one of the zero crossings of the previous $(N-1)/\Delta f_s$.

As can be seen from the last expression in Eq.(2), in the equivalent frequency domain, the sinc pulse sequence is a frequency comb of bandwidth $\Delta f_s$:

$$\text{Comb}_{N\Delta f_s}(f) = \left[\mathbf{F}_t\left(\text{sq}_{N\Delta f_s}(t)\right)\right](f) = \frac{1}{N}\sum_{k=-\infty}^{\infty} \delta\left(f - \frac{k\Delta f_s}{N}\right) \qquad (3)$$

with $N$ as the number and $\Delta f_s/N$ as the frequency distance between the comb lines [17].

Thus, a sinc pulse sequence is a flat, rectangular frequency comb with phase-locked frequencies. Therefore, in contrast to a single sinc pulse, high-quality sinc pulse sequences can be generated by appropriately biased intensity modulators, for instance [17, 18].

Ideal sinc-pulse sequences, with the bandwidth $\Delta f_\mathrm{s}$ can generate ideal signals with the bandwidth $\Delta f_\mathrm{s}/2$. The only difference to the single sinc pulse is, that not an unlimited number of pulses is required (one pulse for each sampling point). Instead, only $N$ time-shifted sequences are needed. Additionally, each single sequence is weighted with periodic sampling points. This periodicity is defined by the number of comb lines $N$ and means that for each of the $N$ branches of the DAC the sampling rate and the required bandwidth is reduced by $N$.

For $N = 3$ and an intensity modulator or an electrical multiplication with the bandwidth $\Delta f_\mathrm{M}$ the sampling rate of the DAC would be $\Delta f_\mathrm{s} = 3\Delta f_\mathrm{M}$ and the baseband bandwidth of the signal is $1.5\,\Delta f_\mathrm{M}$. Thus, with 100 GHz modulators, even available on an integrated platform [19], sampling rates of 300 GS/s and analog bandwidths of 150 GHz could be achieved. However, it has been shown that with integrated modulators even 6 times the bandwidth might be feasible [20], suggesting the generation of signals with analog bandwidths of 300 GHz for integrated DAC.

The basic principle of the orthogonal sampling based DAC is shown in Fig. 1. For the sake of simplicity, the number of branches was restricted to $N = 3$, but we will discuss the principle for an arbitrary number of $N$. The discrete data signal in each of the $N$ branches is:

$$s_l = \frac{N}{\Delta f_\mathrm{s}} \sum_{k=-\infty}^{\infty} s\left(\frac{l-1}{\Delta f_\mathrm{s}} + \frac{Nk}{\Delta f_\mathrm{s}}\right) \cdot \delta\left(t - \frac{l-1}{\Delta f_\mathrm{s}} - \frac{Nk}{\Delta f_\mathrm{s}}\right)$$

(4)

with $l$ as the actual number of the $N$ possible branches. For incrementing $l$, the orthogonality between the sinc pulse sequences is ensured by $1/\Delta f_\mathrm{s}$. The sampling rate of the data for each branch is $\Delta f_\mathrm{s}/N$ (see Fig.1 (a)). If the Dirac Delta sequence of data points is filtered by a rectangular filter with a bandwidth $\Delta f_\mathrm{s}/(2N)$ in each branch, it corresponds to an ideal signal with the same bandwidth, Fig.1 (b). Thus, for generating a signal with the bandwidth $\Delta f_\mathrm{s}/2$ with an $N$-branch system, electronics with a sampling rate of $\Delta f_\mathrm{s}/N$ and a bandwidth of $\Delta f_\mathrm{s}/(2N)$ is required.

In a next step the sampling points are multiplied with a sinc pulse sequence. For $N = 3$ the sinc pulse sequence has two zero crossings. Such a sinc pulse sequence is just a single frequency with an additional direct current (DC) in the frequency domain, or a DC shifted sinusoidal in time, as shown in the insets (c) in Fig.1. For an arbitrary number of $N$, it is a DC together with $n = (N$-1)/2 RF frequencies.

The time-shift between the sequences is a phase shift of the sinusoidals. Therefore, the output of the oscillator is divided into $N$ branches and in each branch the sinusoidal is phase shifted by $360°/l$ ($0°$, $120°$ and $240°$ for $l = 1$, 2 and 3). Afterwards, the sinc pulse sequences are multiplied with the data points in an electrical multiplier or in an optical intensity modulator. In the last step the orthogonal sinc pulse sequences, weighted with the periodic sampling points are added up together to build the signal with $N$-times the bandwidth and sampling rate. This can be described by:

$$s(t) = \sum_{l=1}^{N} \left[ s_l * \left[ \mathbf{F}_t^{-1}\left( \mathrm{Rect}\left(\frac{Nf}{\Delta f_\mathrm{s}}\right) \right) \right] \right](t) \cdot \mathrm{sq}_{N,\Delta f_\mathrm{s}}\left(t - \frac{l-1}{\Delta f_\mathrm{s}}\right)$$

$$= \sum_{l=1}^{N} \underbrace{\sum_{k=-\infty}^{\infty} s\left(\frac{l-1}{\Delta f_\mathrm{s}} + \frac{Nk}{\Delta f_\mathrm{s}}\right) \cdot \mathrm{sinc}\left(\frac{\Delta f_\mathrm{s}}{N} t - \frac{l-1}{N} - k\right)}_{l-th.\ analog\ signal}$$

$$\cdot \underbrace{\mathrm{sq}_{N,\Delta f_\mathrm{s}}\left(t - \frac{l-1}{\Delta f_\mathrm{s}}\right)}_{l-th.\ sinc\ pulse\ sequence}$$

(5)

with Rect as a rectangular function, which has an overall height and width of 1. Each of the $N$ sinc pulse sequences exactly samples the data points from one of the slow rate sub-signals in its peak positions. Due to the orthogonality of the sequences, there is no mutual influence between the time-interleaved data points. In principle, Eq. (5) can be seen as a variant of the sampling theorem (Eq. (1)), where single sinc pulses are replaced by sinc pulse sequences.

## III. SIMULATION AND EXPERIMENTAL RESULTS

For the proof of concept, simulations and experiments were carried out. In additional simulations we have investigated the performance of the method in terms of the jitter and noise influence.

### A. SIMULATION RESULTS

The simulation was carried out with the Optisystem software from Optiwave. Figure 2 shows the simulation setup for the generation of I-Q data signals in three branches, which follows the basic principle of Fig.1. Analog signals can be generated with one of the dashed boxes.

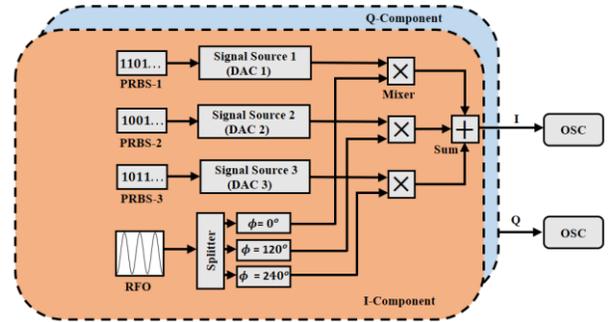

Fig. 2 Simulation setup to generate broad-bandwidth signals from 3 low-bandwidth DACs. PRBS: pseudo random bit sequence, Sum: electrical summator, and RFO: radio frequency oscillator.

First we have generated a 30 GHz sinusoidal signal (Fig. 3(a)) with three 10 GHz DAC by the proposed method (red) and compare this with the direct generation of a 30 GHz signal in a 30 GHz DAC (blue). The root mean square error (rms) was determined as 0.43% for 550,000 samples when calculating the difference between the two signals. For the simulations presented in Fig.3 and 4, the low and high-bandwidth DAC were assumed as ideal. So, they do not have any jitter or noise. In Fig. 3(b) the three low bandwidth signals used to generate the high-bandwidth one can be seen as dashed lines.

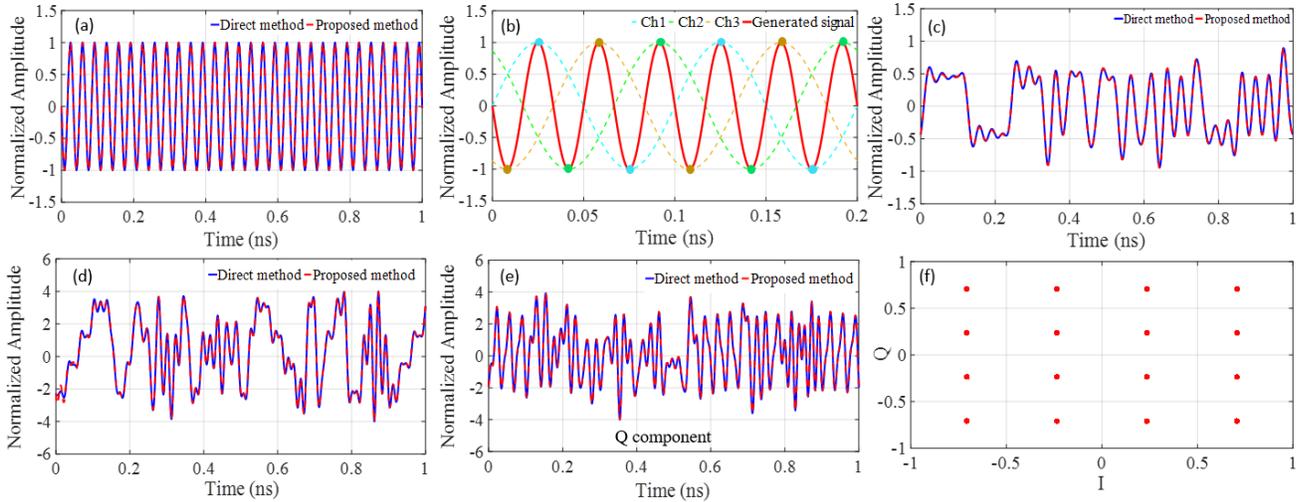

Fig. 3 Simulation results for the generation of a 30 GHz analog signal (a) with a 30 GHz ideal DAC (blue) and three 10 GHz ideal DACs with the proposed method (red). The three analog 10 GHz signals in the sub-branches are presented as dashed lines in (b) together with the generated signal (solid). (c) Generation of a 60 Gbit/s BPSK signal limited to its Nyquist shape (30 GHz bandwidth) by the direct (blue) and proposed method (red). The generation of a 480 Gbit/s, QAM-16 signal limited to its Nyquist bandwidth of 60 GHz for I and Q is shown in (d) and (e). Here again the blue trace shows the direct generation of the signal, whereas the red one presents the signal generation with three 20 GHz sub-DACs. The constellation diagram is shown in (f).

In a second simulation a 60 Gbit/s BPSK signal limited to its Nyquist bandwidth of 30 GHz was simulated. Each sub-DAC with 10 GHz bandwidth and 20 GS/s sampling rate generates a Nyquist sub-signal from the pseudo random bit sequence (PRBS) data signal. Afterwards, these sub-signals are multiplied with a DC shifted, 20 GHz radio frequency. All three sub-signals are added up orthogonally to generate the 60 Gbit/s BPSK Nyquist signal in Fig. 3(c). The rms error between the proposed method and the direct generation by a high-bandwidth DAC was 0.69% for 550,000 samples.

In a next step we have generated a 480 Gbit/s Nyquist QAM-16 signal with three sub-DACs with a bandwidth of 20 GHz. This has been done for I and Q as shown in Fig.2. The signals in each of the three branches are multiplied with a DC shifted 40 GHz RF tone. The results for I and Q are shown with the red trace in Fig. 3(d), and (e), respectively. The black trace shows the result for a 60 GHz ideal DAC for comparison. The calculated rms error was approximately 2% for 550,000 samples. The constellation diagram for the proposed method is shown in Fig. 3(f).

A further bandwidth reduction for the electronics in each sub branch can be achieved by increasing the number of parallel branches $N$. Therefore, in further simulations we have simulated the generation of the same signals as in Fig.3 but, with five parallel sub-branches. In this case, the single DAC in the sub-branch only needs a bandwidth of 6 GHz and the oscillator has to provide two RF frequencies of 12 and 24 GHz. The orthogonality between the 5 sinc-pulse sequences can be obtained by adjusting the phases between the RF to 0º, 72º, 144º, 216º, and 288º. As illustrated in Fig. 4(a), the 30 GHz signal shows an rms error of 0.43% for 550,000 samples. Figure 4(c) presents the 60 Gbit/s BPSK signal, generated by 5 sub-DACS with 6 GHz bandwidth.

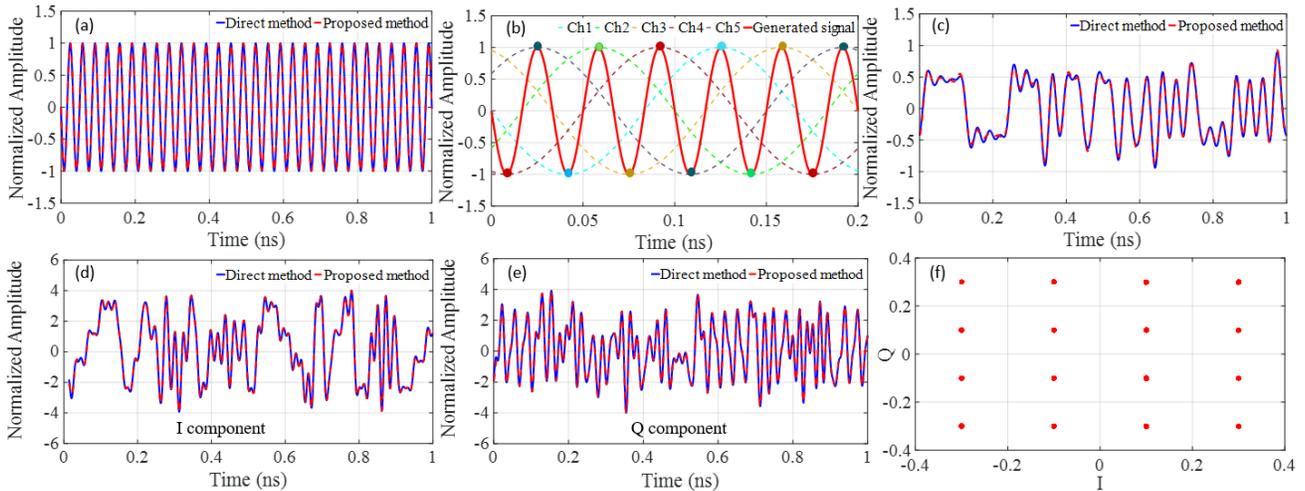

Fig. 4. Generation of a 30 GHz analog signal (a, b) and a 60 Gbit/s BPSK Nyquist signal (c) with 5 sub-DACs with a bandwidth of 6 GHz. (d − f) generation of a 480 Gbit/s QAM-16 Nyquist signal with 60 GHz bandwidth by 5 sub-DACs with a bandwidth of 12 GHz.

Compared to the direct generation of the signal, the rms error was 1.1 % for 550,000 samples. In Fig. 4 (d − f) the generation of a 480 Gbit/s QAM-16 Nyquist signal with 5 parallel sub-DACs with a bandwidth of 12 GHz is presented. The measured rms error to the direct generation with an ideal 60 GHz DAC was 2% for 400,000 samples.

As mentioned, in the above simulations ideal DACs without jitter and noise were assumed. In order to evaluate the performance of the proposed method, an effective number of bit (ENOB) for the generation of the 30 GHz analog signal with 3 and 5 sub-DACs was conducted. Please note that for the proposed method there are two jitter values, i.e. the jitter of the RF oscillator generating the sinusoidal waves and the jitter of the electronic sub-DACs. The jitter of the RF oscillator can be extremely small in the zepto second range [21] and even for integrated oscillators jitter values of 20 fs have been shown [22]. In the simulation, however, we have assumed the jitter for the RF source as 100 fs. The ENOB as a function of the jitter is shown in Fig. 5. As can be seen, for the same jitter value the proposed method with three (red) or five branches (black) not only drastically reduces the bandwidth requirements on the electronics, at the same time the quality of the generated signal is enhanced. For a jitter of 1 ps, for instance, the ENOB is increased by 2. This corresponds to a signal-to-noise-and-distortion (SINAD) enhancement of around 12 dB.

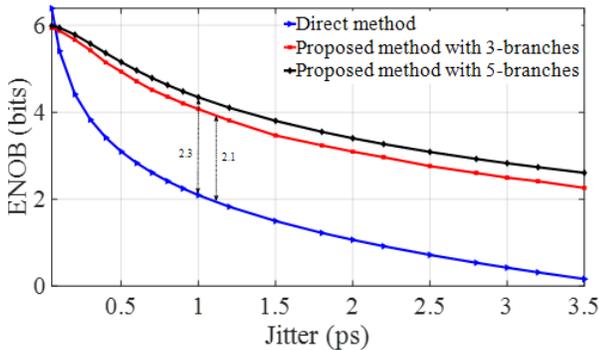

Fig. 5. ENOB vs. jitter simulation for the generation of a 30 GHz sinusoidal signal with the proposed method in 3 and 5 parallel branches in comparison to a direct generation.

To show the performance enhancement which can be achieved by the proposed method for communication applications, we have simulated an optical link as shown in Fig.6. The Mach-Zehnder modulator (MZM) converts the signal generated with a high-bandwidth DAC (a) or by the proposed method with 3 or 5 sub-DACs (b) into the optical domain at a frequency of 193.1 THz (C-band of optical telecommunications). Optical noise [provided by an optical signal-to-noise-ratio module (OSNR) of the software] was added to the signal. A coherent detector (CD) with a 30 GHz bandwidth was used to detect the signal.

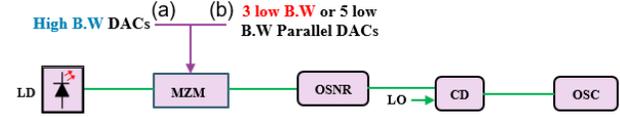

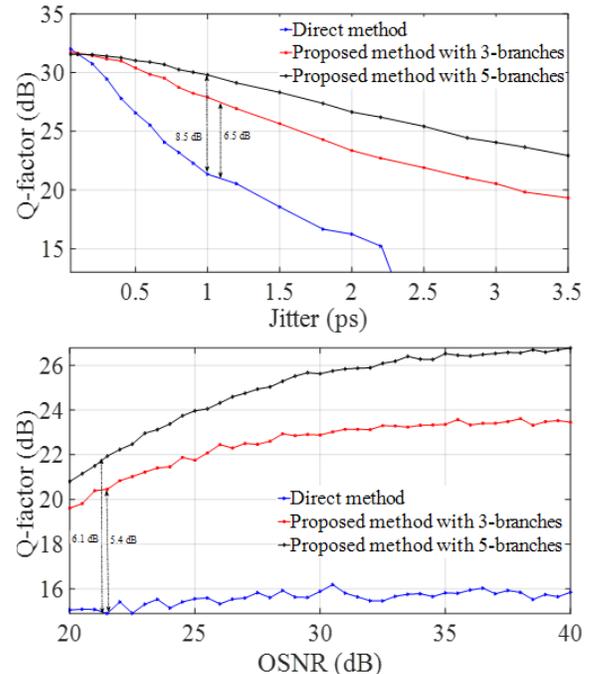

Fig. 6. Simulated optical link for the transmission of a 60 GBd BPSK signal limited to the Nyquist shape of 30 GHz from a DAC with 30 GHz bandwidth (a) and three or five sub-DACs with 10 GHz or 6 GHz bandwidth (b). LD: laser diode, MZM: Mach-Zehnder modulator, OSNR: optical-signal-to-noise-ratio component, LO: local oscillator, CD: coherent detector, and OSC: electrical oscilloscope.

The Q-factor vs jitter for the transmission of a 60 GBd BPSK Nyquist signal, generating the signals with a 30 GHz DAC (blue) or with three or five sub-DACs with 10 or 6 GHz bandwidth (red and blue), are shown in Fig.7 (a). Again, the jitter of the RF oscillator was assumed as 100 fs, while the jitter of the electronic DAC was swept from 0.05 ps to 3.5 ps. The OSNR was 40 dB. As can be seen, the higher SINAD of the generated signals is transferred to the transmission. For a jitter of 1 ps the Q-factor of the received signal improves by around 6.5 and 8.5 dB if the transmitted signal is generated in three or five branches, respectively. For electronic DAC with a jitter above 2.2 ps the link collapses. However, with the presented method even such low-quality electronic DAC can still be used for a transmission system.

The effect of noise is presented in Fig. 7(b). The OSNR was swept from 20 to 40 dB and the jitter was assumed as 100 fs for the RF source and the electronic DACs. As shown, the Q-factor for the transmission of the directly generated signals is almost constant over the range of OSNR presented in the figure. For the higher quality signals generated in 3 or 5 branches, the Q-factor of the received signal is improved.

Fig. 7. Performance simulation of the proposed method with 3 and 5 branches for an optical link in comparison to the direct generation of the signals. In (a) the jitter of the RF oscillator was kept constant at 100 fs and for the DACs it was swept. In (b) the OSNR performance for a 100 fs jitter of the RF source and the electronic sub-DACs is depicted.

*B. EXPERIMENTAL RESULTS*

In proof of concept experiments, different types of high-bandwidth signals with 3 and 5 parallel sub-DACs were generated and compared to directly generated ones. Unfortunately, electrical multipliers and adders were not available. Therefore, the signals were created, as described in Fig.1, by Mathematica and then provided to an arbitrary waveform generator (AWG) (Tektronix AWG70001A). The generated signals were then measured with an electrical sampling oscilloscope (Tektronix DP073304SX).

Since we did not use parallel low bandwidth DACs but just one single high bandwidth one, the improvement of the signal generation by the proposed method cannot be seen. However, as presented in Fig. 8, the analog 10 GHz sinusoidal signal as well as the 24 Gbit/s Nyquist or 12 Gbit/s normal data signal are almost the same.

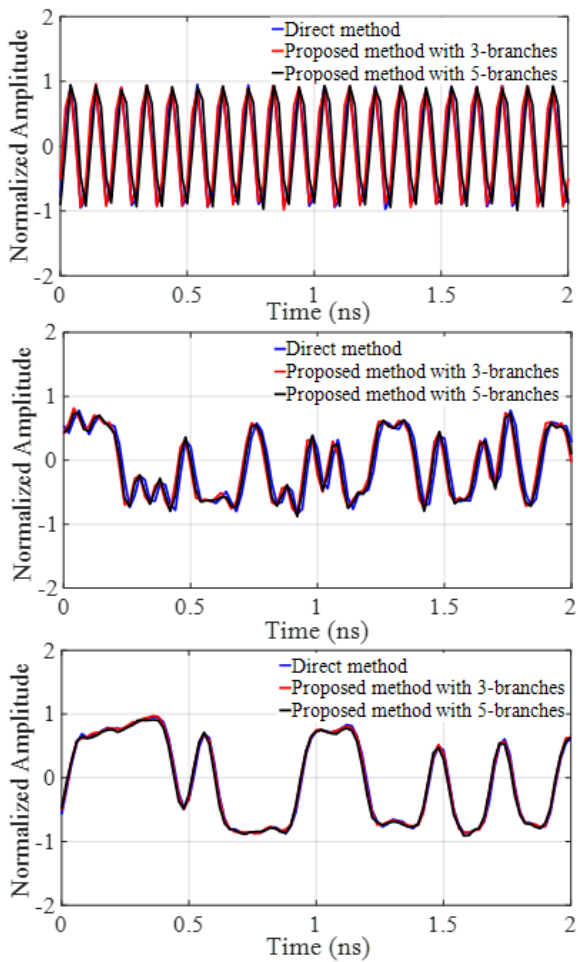

Fig. 8. Experimental results for the generation of (a)10 GHz analog signal, (b) 24 Gbit/s BPSK Nyquist signal, and (c) 12 Gbit/s BPSK normal signal with the direct method (blue) and with 3 (red) and 5 parallel branches (black).

## IV. CONCLUSION

In conclusion, a new concept for generating wide-bandwidth signals with low-bandwidth electronics was presented. The proposed method is based on orthogonal sampling with sinc-pulse sequences in parallel sub-branches. The sampling rate and bandwidth of the generated signals is increased linearly with the number of parallel branches $N$. As shown, not only the bandwidth requirements on the electronics are drastically reduced, at the same time the generated signals show a higher ENOB and SINAD, which can be directly transferred to the transmission quality in communication systems. The method only requires an RF oscillator, electrical multipliers and an adder. Thus, the integration might be straight forward. We believe that the method is important for all broad-bandwidth applications.

## ACKNOWLEDGEMENT

The authors would like to acknowledge the financial support of the German Research Foundation, Deutsche Forschungsgemeinschaft, under grant numbers -403154102, 322402243, 424608109, 424608271, 424607946, 424608191, 491066027- and in part by the German Federal Ministry of Education and Research (BMBF), under Grant 13N14879.